\documentclass[letterpaper, 10 pt, journal, twoside]{IEEEtran}% Comment this line out if you need a4paper

\IEEEoverridecommandlockouts                              % This command is only needed if 
\usepackage{booktabs}
\let\labelindent\relax
\usepackage{enumitem}

\usepackage{xcolor}
\definecolor{dark-green}{RGB}{12,80,12}

\newcommand{\expnumber}[2]{{#1}\mathrm{e}{#2}}

\usepackage[pagebackref,breaklinks,colorlinks]{hyperref}
\usepackage{siunitx}
\usepackage[ruled,vlined]{algorithm2e}
\usepackage{algorithmic}
\usepackage{microtype}
\SetAlFnt{\small}
\SetAlCapFnt{\small}
\SetAlCapNameFnt{\small}

\usepackage[capitalize]{cleveref}
\crefname{section}{Sec.}{Secs.}
\Crefname{section}{Section}{Sections}
\Crefname{table}{Table}{Tables}
\crefname{table}{Tab.}{Tabs.}
\crefname{algorithm}{Algo.}{Algos.}

\newcommand{\tablesize}{\fontsize{8pt}{10pt}\selectfont}

\newcommand{\website}{\url{http://dav-nav.cs.uni-freiburg.de}}

\usepackage{graphicx}
\usepackage{amsfonts} % assumes amsmath package installed

\usepackage[font=small]{caption}
\usepackage{cuted}

\newcommand{\myworries}[1]{\textcolor{black}{#1}}
\newcommand*{\rowstyle}[1]{% sets the style of the next row
  \gdef\@rowstyle{#1}%
  \@rowstyle\ignorespaces%
}
\newcolumntype{=}{% resets the row style
  >{\gdef\@rowstyle{}}%
}
\newcolumntype{+}{% adds the current row style to the next column
  >{\@rowstyle}%
}

\title{%\LARGE \bf
Catch Me If You Hear Me: Audio-Visual Navigation in Complex Unmapped Environments with Moving Sounds
}

\author{Abdelrahman Younes$^*$, Daniel Honerkamp$^*$, Tim Welschehold, and Abhinav Valada% <-this % stops a space

\thanks{© 2022 IEEE.  Personal use of this material is permitted.  Permission from IEEE must be obtained for all other uses, in any current or future media, including reprinting/republishing this material for advertising or promotional purposes, creating new collective works, for resale or redistribution to servers or lists, or reuse of any copyrighted component of this work in other works.}
\thanks{This work was supported by the European Union’s Horizon 2020 research and innovation program under grant
agreement No 871449-OpenDR.)} %Use only for final RAL version
\thanks{$^*$These authors contributed equally. All authors are with the Department of Computer Science, University of Freiburg, Germany.}%
}

\begin{document}

\maketitle

\begin{abstract}
Audio-visual navigation combines sight and hearing to navigate to a sound-emitting source in an unmapped environment. While recent approaches have demonstrated the benefits of audio input to detect and find the goal, they focus on clean and static sound sources and struggle to generalize to unheard sounds. In this work, we propose the novel dynamic audio-visual navigation benchmark which requires catching a moving sound source in an environment with noisy and distracting sounds, posing a range of new challenges. We introduce a reinforcement learning approach that learns a robust navigation policy for these complex settings. To achieve this, we propose an architecture that fuses audio-visual information in the spatial feature space to learn correlations of geometric information inherent in both local maps and audio signals. We demonstrate that our approach consistently outperforms the current state-of-the-art by a large margin across all tasks of moving sounds, unheard sounds, and noisy environments, on two challenging 3D scanned real-world environments, namely Matterport3D and Replica. The benchmark is available at \website.
\end{abstract}
% Keywords appear just beneath the abstract. Use only for final RAL version. 
\begin{IEEEkeywords}
 Autonomous Agents, Reinforcement Learning, Reactive and Sensor-Based Planning
\end{IEEEkeywords}
% \begin{IEEEkeywords}Embodied AI, Audio-Visual Navigation, Reinforcement Learning
% \end{IEEEkeywords}

\section{Introduction}\label{sec:intro}
% \blfootnote{$^*$Equal contribution.}
\IEEEPARstart{H}{umans}
are able to very efficiently combine their senses of hearing and sight in order to navigate unknown environments. While navigation in such environments has been an important focus of embodied AI~\cite{gupta2017cognitive,wijmans2019dd}, existing work on navigation overwhelmingly relies on sensors such as vision and LiDAR, leaving out other core senses used by humans. Sound is a particularly unique modality as it reveals information beyond the visible walls and obstacles~\cite{valada2017deep}. In particular, it provides blind people with spatial navigation capability comparable to sighted people~\cite{fortin2008wayfinding}.

Recent work has demonstrated the value of this signal for embodied agents in a variety of tasks. This includes audio-visual navigation, in which the agent is required to navigate to the location of a sound-emitting source using audio and visual signals~\cite{chen2020soundspaces,gan2020look}, semantic audio-visual navigation~\cite{chen2021semantic} with coherent room and sound semantics, active perception tasks such as active audio-visual source separation \cite{majumder2021move2hear} and audio-visual dereverberation~\cite{chen2021learning}, curiosity-based exploration via audio-visual association~\cite{valverde2021there} as well as tasks explicitly focusing on the geometric information contained in audio such as audio-visual floor plan reconstruction~\cite{purushwalkam2020audio,boniardi2019robot}.

\begin{figure}
    \centering
    \includegraphics[width=0.399\linewidth,trim={3.4cm 2.0cm 6.1cm 4.2cm},clip]{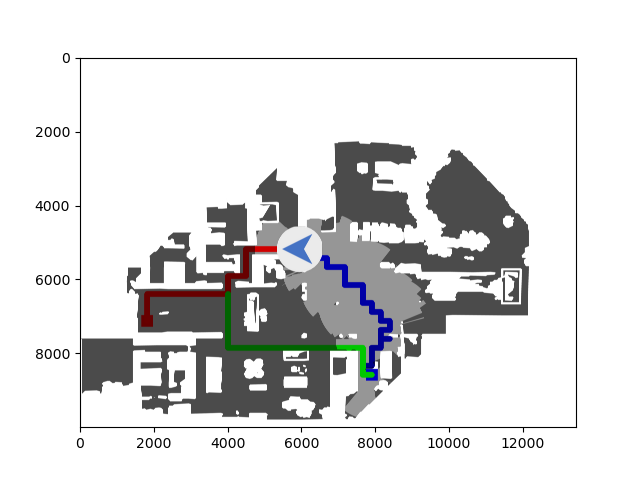}
    \includegraphics[width=0.5453\linewidth,trim={0.8cm 0.5cm .9cm 2.1cm},clip]{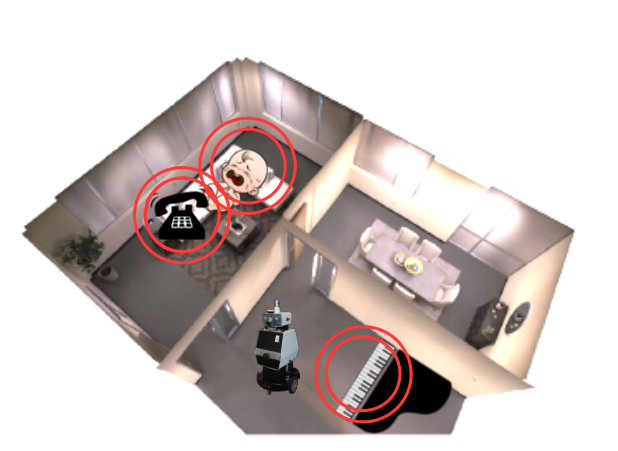}
    \caption{We introduce a novel dynamic audio-visual navigation benchmark (left). The paths of the agent and the sound source are shown in blue and red respectively, with initial poses marked as squares. The green line represents the optimal behavior to catch the moving target. Secondly, we carefully design complex audio scenarios (right). The agent needs to navigate towards the ringing phone while being confronted with a second sound source (a crying baby), and various distractor sounds such as a piano.}
    \label{fig:paper-teaser}
    %\vspace{-0.5cm}
\end{figure}

Navigation-centered approaches have shown that agents can successfully extract information from audio signals. However, they have mostly focused on clean and distractor-free audio settings in which the only change to the audio signal comes from changes in the agent's position. Furthermore, they have struggled to generalize to unheard sounds~\cite{chen2020soundspaces,chen2020learning}. 
In this work, we take the next steps toward more challenging scenarios. % by lifting most of prior work assumptions. 
First, we introduce a novel dynamic audio-visual navigation benchmark with a moving sound source. This extends the applicability by capturing common scenarios such as a robot navigating to a person issuing commands or following pets or people in the house. We argue that this strongly increases the complexity of the task through two channels: on one hand, previous observations no longer capture the current state of the environment and the agent has to learn to update its memory accordingly. On the other hand, optimal behavior now requires not just following the sound intensity but proactively reasoning about the movement of the target to catch it efficiently.
Secondly, we carefully design complex audio scenarios with augmented, noisy and distracting sounds that lift many assumptions about clean audio signals made by previous work and show the benefits of training on these scenarios for generalization to unheard sounds. \cref{fig:paper-teaser} illustrates both the moving sound task (left) and the complex audio scenarios we construct (right). Lastly, we introduce an architecture that explicitly enables the agent to spatially fuse the geometric information inherent in obstacle maps and audio signals. We show that this leads to significant gains in the generalization to unheard sounds in both clean and complex audio scenarios. We demonstrate these results in the SoundSpaces~\cite{chen2020soundspaces} extension to the Habitat simulator~\cite{habitat19iccv}, which allows us to generate realistic binaural sound signals for the realistic 3D environments of the Replica~\cite{straub2019replica} and Matterport3D~\cite{chang2017matterport3d} datasets. Combined, these contributions achieve improvements over the previous state-of-the-art by 36\% and 29\% in the success rate as well as an improvement in SPL by 37\% and 37.1\% on Replica and Matterport3D respectively, on the unheard and static AudioGoal benchmark~\cite{chen2020soundspaces}. An early version of the approach won first place in the 2021 CVPR SoundSpaces Challenge\footnote{\url{https://soundspaces.org/challenge}.}.

To summarise, the main contributions of this work are:
\begin{itemize}[noitemsep, topsep=0pt]
    \item We introduce the novel dynamic audio-visual navigation benchmark that generalizes previous audio-visual navigation tasks and poses new challenges.
    \item We propose complex audio scenarios with noisy and distracting sounds and demonstrate the benefits of this randomization for better generalization.
    \item We propose a new architecture that allows spatially fusing sound and vision, and  outperforms current state-of-the-art approaches by up to 37\% on unheard sounds.
    \item We perform exhaustive experiments in two realistic 3D environments, and across static and dynamic audio goal tasks in both clean and complex audio scenarios.
    % \item The code and benchmark are publicly available at \website.
\end{itemize}

% \    \item Heard sounds

% \comm{
% Challenges in these tasks:
% \begin{itemize}
%     \item Integrate sound over time to filter out noise and distractors
%     \item RL part: move ideally to sample the observations that tell us best where the sound is
%     \item Termination: accurately recognise when to stop
%     \item Generalisation to unheard sounds and unseen apartments
%     \item Moving sounds: don't just move towards the sound source, but move to 'catch' it
%     \item Top-down map: includes walls as well as tables etc. Sound cannot travel through walls, but it can travel over tables. If target is in a corner, the navigation policy might have to travel in a different direction than the sound for a prolonged time
% \end{itemize}
% Our contributions
% \begin{itemize}
%     \item Introduce active audio tracking with moving sound source
%     \item Propose an auxiliary task to reconstruct the binaural spectogram
%     \item Introduce a novel architecture for the learning agent
%     \item 
% \end{itemize}
% }

\section{Related Work}\label{sec:relatedwork}
{\parskip=5pt
\noindent\textit{Embodied Navigation}: The ability to navigate the physical world has been a core focus of research in embodied AI. Much progress has been achieved on tasks such as PointGoal Navigation~\cite{gupta2017cognitive,chen2019behavioral,wijmans2019dd}, ObjectGoal Navigation~\cite{yang2018visual,du2021vtnet}, Vision-Language Navigation~\cite{fried2018speaker,chen2021topological}, Active Visual Tracking~\cite{zhong2021towards,hurtado2020mopt}, and Visual Exploration~\cite{pathak2017curiosity,chaplot2020learning}.
While most common goals are assumed to be static, some work has been conducted on moving goals.
Control approaches to track and intercept moving sounds have been proposed based on explicit forecasting of the target's movement, but rely on first visually locating the target~\cite{belkhouche2007,ZHU2013}.
Work in multi-agent systems and social navigation~\cite{chen2017decentralized,forer2018socially,hurtado2021learning} investigate scenarios in which multiple agents influence each other, both collaboratively and competitively. In contrast, we focus on cases in which the target is frequently outside the field of visual sensors and assume that the agent's behavior does not impact the sound source's movements.}

{\parskip=5pt
\noindent\textit{Sound Source Localization}: The localization of sound sources based on an audio signal has been explored in robotics for both static~\cite{nakadai1999sound,rascon2017} and dynamic sounds~\cite{valin2004localization,evers2015bearing}. Early work in combining audio and vision focused on using Gaussian Processes~\cite{hershey1999audio} or canonical correlation analysis~\cite{kidron2005}. Recently, deep learning based methods have been proposed to localize sounding objects in videos~\cite{tian2018audio,Arandjelovic_2018_ECCV}. In contrast, audio-visual navigation problems focus on potentially out-of-sight sound sources in unmapped environments.}

{\parskip=5pt
\noindent\textit{Audio-Visual Navigation}: Recent simulators and datasets have enabled the training of learning-based systems on the combination of visually realistic scenes and varied audio signals~\cite{chen2020soundspaces,gan2020look}.
In the AudioGoal task, an agent navigates to a continuously sound-emitting target source's location using audio and visual signals~\cite{chen2020soundspaces,gan2020look}. This task has been tackled by decomposing the problem into the prediction of the sound location and a planner~\cite{gan2020look} or by an end-to-end reinforcement learning approach with either low-level actions \cite{chen2020soundspaces} or on a higher level, combining learned waypoints and a planner~\cite{chen2020learning}. However, not much focus has been put on the fusion of sound and visual observations.
%relying on standard RNN cells to merge the individually encoded and concatenated features. 
In contrast, we provide the agent with an architectural prior that allows it to explicitly learn to fuse spatial information from both modalities.
Semantic Audio-Visual Navigation~\cite{chen2021semantic} introduces a related task in which the sound is periodic and semantically consistent with the scene, allowing agents to exploit semantic understanding between sound and vision to navigate towards the goal. 
However, these tasks assume a single static sound-emitting source. While this setup covers a large number of potential use cases, it is still a subset of the scenarios that humans navigate. We propose a novel task where the agent tracks a moving sound source using only acoustic and visual observations.}
% Audio-Visual Floor Plan Reconstruction~\cite{purushwalkam2020audio} uses audio as an additional signal to the visual input to enrich the perception about the surrounding environment to reconstruct the floor plan.

Previous work focuses either solely on clean audio scenarios \cite{chen2020soundspaces,gan2020look} or provides an initial but limited exploration of more complex scenarios. This includes the presence of a distractor or microphone noise on the single heard AudioGoal task~\cite{chen2020learning} and evaluation of the impact of distractors in related tasks such as Semantic Audio-Visual Navigation~\cite{chen2021semantic} and Active Audio-Visual Source Separation~\cite{majumder2021move2hear}, in which the agent needs to move in an intelligent manner to separate the input target monaural sound source from distractor sounds. \myworries{SAAVN~\cite{yu2022sound} trains an adversary to distract the agent with a sound signal. But is limited to heard sounds and distractors and a single adversary moving step-by-step around the apartment.}
In contrast, we extensively evaluate the impact of both training and testing in the presence of a variety of strong audio perturbations in audio-visual navigation and demonstrate the strong benefits of these scenarios for the generalization to unheard sounds \myworries{and unheard distractors}. Unlike in audio source separation, the agent does not get access to a clean target sound.
% \myworries{SAAVN~\cite{yu2022sound} trains an adversary to distract the agent with a sound signal. But is limited to heard sounds and distractors and a single adversary moving step-by-step around the apartment. In contrast, we focus on generalization to both unheard target and unheard distractor sounds and general audio perturbations.}

{\parskip=5pt
\noindent\textit{Augmentation and Domain Randomisation}: Data augmentation~\cite{shorten2019survey} and domain randomization~\cite{fereshteh2017cad2rl,james2017transferring} have shown to be very beneficial in regimes of limited data, to improve generalization to unseen data and to obtain robustness against noise. Audio-specific augmentations include transformations such as time warping, frequency masking, and time masking~\cite{park2019specaugment}. In this work, we introduce complex audio scenarios specific to audio-visual navigation that similarly increases the variety and diversity of the training data.}

{\parskip=5pt
\noindent\textit{Active Visual Tracking}: Distractors have been used within the visual domain, learning to keep tracking objects or people of interest within the field of view \cite{zhong2021towards}. In contrast, we present a setting of audio-distractors and randomizations.}

%\section{Technical Approach}
\section{Problem Statement}
We tackle the challenge of navigating to a sound-emitting goal.
We use the highly photo-realistic datasets Replica~\cite{straub2019replica} and Matterport3D~\cite{chang2017matterport3d}, which consist of 3D scans of indoor areas such as homes, offices, hotels, and rooms. Replica scenes' areas vary between $\SI{9.5}{\meter}$ to $\SI{141.5}{\meter\squared}$ while Matterport3D areas vary between $\SI{53.1}{\meter}$ to $\SI{2921.3}{\meter\squared}$ providing the agent with diverse experience of different real-world scenes with close and far AudioGoals. We use Habitat~\cite{habitat19iccv} and its audio-compatible SoundSpaces simulator~\cite{chen2020soundspaces} to train our agents. The SoundSpaces simulator offers binaural room impulse response (BRIR) between every two possible source and receiver locations on a grid with a spatial resolution of $\SI{0.5}{\meter}$ for Replica and $\SI{1}{\meter}$ for Matterport3D dataset. The pre-calculated BRIR can be convolved with any arbitrary sound to simulate how the receiver listens to this audio at the agent's current location. We use the same 102 copyright-free sounds used in the AudioGoal benchmark~\cite{chen2020soundspaces}, available under CC-BY-4.0 licence.
The BRIRs have a sampling rate of \SI{16000}{\hertz} for Matterport3D and \SI{44100}{\hertz} for Replica. The spectrograms are computed as in \cite{chen2020soundspaces}, \cite{chen2020learning}: We compute a Short-Time Fourier Transform (STFT) with a window length of 512 samples and a hop length of 160. From this, we take the magnitude, downsample the axes by four, and compute the logarithm. Stacking the left and right audio channels then results in a spectrogram of size (65,26,2) for Matterport3D and  (65,69,2) for Replica.

We formulate the problem as a reinforcement learning task where the agent learns to navigate an unknown environment to reach a potentially previously unheard sound-emitting goal. In each step, the agent receives the current observation $o_t$ consisting of RGB image $v_t$ and depth image $d_t$ as well as a binaural sound $b_t$ in the form of a spectrogram for the right and left ear. In contrast to the common PointGoal navigation task, the agent does not receive a displacement vector to indicate the goal position. 
Given the current observation and the agent's previous state $s_{t-1}$, the agent then produces a next action $a_t$ from its policy $\pi(a_t | o_t, s_{t-1})$. The agent's aim is to maximise the expected discounted return $\mathbb{E}_\pi[\sum_{t=1}^{T} \gamma^t r(s_{t-1}, a_t)]$, where $\gamma$ is the discount factor and $r(s_{t-1}, a_t)$ is the reward.
Note that the sound goal does not have an embodiment, i.e. it is not visible in the RGB-D images and cannot collide with the agent.

We build upon the AudioGoal task as introduced by \cite{chen2020soundspaces}. The agent starts in a random pose in an unknown environment. It then has to navigate to the sound location and execute the \textit{Stop} action at the exact position of the sound source. 
The discrete action space of the habitat simulator consists of \textit{Move Forward, Rotate Left, Rotate Right}, and \textit{Stop}. %The parameterization of our reinforcement learning agent’s action space, however, follows a waypoint selection approach \cite{chen2020learning}.
To increase the capabilities and possible uses of such an agent, we extend the task to moving noise-emitting targets. We further seek to improve the agent’s performance in general and in particular with regard to navigating to previously unheard sounds.

\section{Technical Approach}

In order to address the aforementioned challenges, we introduce the novel dynamic audio-visual navigation benchmark which largely increases the demands on both the agent's memory as well as its policy which now has to proactively move and catch the agent to act optimally.
We then introduce complex audio scenarios for both the moving and the existing static AudioGoal navigation task~\cite{chen2020soundspaces}. 
To this end, we develop audio-domain specific scenarios in which the agent is confronted with both episodic and per-step randomizations, requiring integration and filtering the sound signals over time.
We further propose a new architecture that allows the agent to directly integrate the spatial and directional information from sound and vision, and strongly increases robustness and generalization to unheard sounds as we demonstrate in \cref{sec:experiments}.

\subsection{Dynamic Audio-Visual Navigation}
\label{sec:dynamic_task}

We introduce the novel task of dynamic audio-visual navigation. In this task, the agent must navigate towards a moving sound-emitting source in an unmapped complex 3D environment and output \textit{Stop} when it catches it. The agent needs to reason about the trajectory of the moving sound based on audio and visual observations to decide the shortest path to reach it. This can be seen as a generalization of the existing task, broadening the scope of static AudioGoal navigation to scenarios such as navigating to a person issuing commands or following a pet. \myworries{As the target is moving, previous observations on their own no longer point directly to the current position of the target and the agent has to learn to update its memory and aggregate information across time to form an estimate of the target's position and movement}. Furthermore, it is no longer sufficient to follow the gradient of the sound intensity, instead, optimal behavior now requires proactively reasoning about the movement of the target to catch it efficiently. To the best of our knowledge, this is the first approach to investigate the use of sound and vision to catch moving sounds within unexplored environments.

{\parskip=5pt
\noindent\textit{Motion Model}: We assume a simple, goal-direct behavior of the target. The sound source starts in a random pose on the map and uniformly draws a goal from the traversable grid, excluding the agent's current position. We also ensure that there exists a traversable path from its start to the goal location. The sound source then follows the shortest path towards the goal and with a probability of $30\%$ moves to the next node.
% After choosing the target destination, we generate a list containing the intermediate location the moving sound has to pass by while taking the shortest path towards that destination. 
% In every step, the simulator randomize whether to move to the next location in that list or not with probability of $30\%$ to include it in every step. 
This percentage ensures the sound source moves slightly slower than the agent, ensuring that it is possible to catch the moving sound source. Note that the moving source does not have an orientation and directly moves to the following location while the agent has to take separate rotation steps to change direction. We ablate this movement speed in the supplementary material and find that our model is robust to a wide range of values.
Once the moving target reaches its goal, it draws a new random goal to navigate to.}

{\parskip=5pt
\noindent\textit{Optimal Behavior: Dynamic Success weighted by Path Length (DSPL)}: The Success weighted by Path Length (SPL) \cite{anderson2018evaluation} serves as the primary metric to evaluate the navigation performance of embodied agents. However, in the case of a moving sound source, the shortest possible path depends on the a priori unknown trajectory of the sound source. Given this knowledge, the optimal policy is to move to the earliest intersection with the target's trajectory that the agent can reach before the target passes by. Hence, we introduce the Dynamic Success weighted by Path Length (DSPL) to measure how close the agent is to this oracle optimal policy.
% \paragraph{Dynamic Success weighted by Path Length (DSPL)} The Success weighted by Path Length (SPL) \cite{anderson2018evaluation} serves as the primary metric to evaluate the navigation performance of embodied agents. However, this metric relies on the prior knowledge of the shortest path to the goal location, however in our novel task, the goal location changes continually during the episode. Hence, we introduce a modified version of that metric which considers the shortest path length as the geodesic distance between the agent's starting position and the closest position the moving source passed by if the agent had enough steps from the beginning of the episode to catch the source in that position. Therefore, in every step, we calculate the number of steps the agents needs to catch the goal at the current goal position, and if that number is smaller than or equal to the current episode steps count, we consider the current target position as the closest position the agent could catch the source at and we consider the geodesic distance between the agent's start location and current goal location as the shortest distance the agent needs to reach the goal then 
We define the DSPL as follows: where $i$ is the current episode count, $N$ is the total number of episodes, $S_i$ represents whether this episode is successful or not, $g_i$ is the shortest geodesic distance between the agent's start location and the closest position the agent could have caught the sound source at, and $p_i$ is the length of the path taken by the agent:}
\begin{equation}\label{1}
     DSPL = \frac{1}{N}\sum_{i=1}^{N}S_i \frac{g_i}{max(p_i,g_i)}.
\end{equation}

Note that this metric represents an oracle upper bound of the possible performance, which may not be achievable without a priori knowledge of the trajectory of the sound source. An example of the task and the optimal behavior as used in the DSPL is shown in \cref{fig:paper-teaser}.

\subsection{Reward}\label{sec:reward}

We use the same reward definition for both static and dynamic tasks. 
Upon success, the agent receives a positive reward of +10. It further receives a small dense reward of +0.25 for decreasing and -0.25 for increasing the shortest path distance to the goal. 
For the dynamically moving sounds, this is calculated with respect to the current sound source's position and not with regard to the shortest reachable intersection with its trajectory. As a result, in this setting the dense reward no longer directly points to the optimal policy, reducing the value of the supervisory signal. Finally, a small time penalty of -0.01 per step incentivizes the agent to find short paths.\looseness=-1
 
% This increases exploration difficulty but keeps the calculation simple and applicable to real world setups.

\subsection{Complex Audio Scenarios}
\label{sec:complex}
\begin{algorithm}[t]
\tablesize
\textbf{Require: }{targetAudio: current target sound, listOfSounds: training sounds excluding targetAudio, agentPosition, targetPosition, computeSpectrogram, computeAudio: function to compute audio observation, applySpecAugment: function to apply feature augmentation, listOfNodes: traversable grid locations, rnd: uniformly random choice function.}
\newline
\SetAlgoLined
\For{episode in episodes}
{
includeSecondSound = rnd([True, False])

\If{includeSecondSound}{
secondAudio = rnd(listOfSounds)}

includeDistractor = rnd([True, False])

\For{step in steps}{
audio = computeAudio(targetAudio, agentPosition,targetPosition)
\\
\If{includeSecondSound}{
audio += computeAudio(secondAudio, agentPosition,targetPosition)
}
distractorStep = rnd([True, False])

\If{includeDistractor \textbf{and} {distractorStep} }{
distractorAudio = rnd(listOfSounds)
\\
distractorPosition = rnd(listOfNodes)
\\
audio += computeAudio(distractorAudio agentPosition,distractorPosition)
}
spectrogram = computeSpectrogram(audio)
\\
augment = rnd([True, False])
\\
\If{augment}{
aug = rnd([timeMasking, frequencyMasking, both])
\\
applySpecAugment(spectrogram, aug)
}
}
}
 \caption{Randomization Pipeline}
 \label{alg:algorithm1}
\end{algorithm}
\begin{figure*}
  \centering
%   \fbox{\rule{0pt}{2in} \rule{0.9\linewidth}{0pt}}
  \includegraphics[width=.65\linewidth,trim={0cm 0.05cm 0cm 0.3cm},clip]{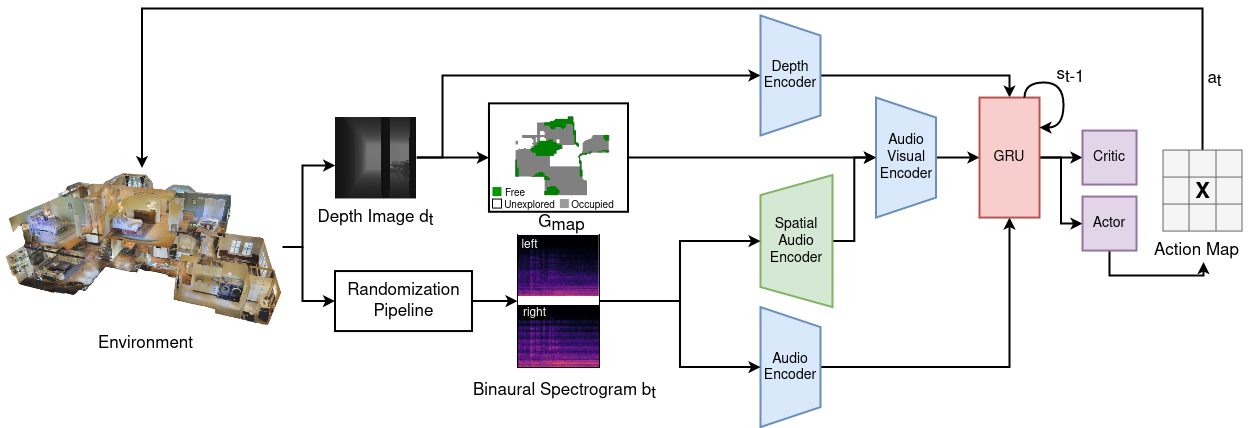}
  \caption{Our proposed architecture. The depth image is projected into an allocentric geometric map $G_{map}$. A novel channel consisting of a Spatial Audio Encoder and Audio-Visual Encoder fuses the spatial information inherent in the geometric map and audio signal. A GRU then combines this channel with depth and audio encodings. A PPO agent produces close-by waypoints that are executed by a Djikstra planner.\looseness=-1}
  \label{fig:architecture}
  %\vspace{-0.3cm}
\end{figure*}

% The ultimate goal of training Embodied AI agents to follow a sound-emitting source is to deploy them in real-life to enhance the quality of people's life by doing some of their tasks that require navigating towards a sound source (e.g., fire alarm). However, 
Current approaches  \cite{chen2020soundspaces,gan2020look,chen2020learning} mainly focus on relatively simple scenarios with a single sound-emitting source in a clean audio environment, while the impact of sound disturbances remains limited, as discussed in \cref{sec:relatedwork}. 
% While \cite{chen2020learning} evaluate their model in the presence of distractors, they only evaluate the case in which the agent has trained on a single sound source and confronted with a disjoint set of distractors, leaving an easier pathway for the agent to simply filter for the single known sound it has been trained on. Our approach is also different from the test conducted on the Semantic Audio-Visual Navigation \cite{chen2021semantic} work, as they added to their input observations a one-hot encoded vector representing the sound the agents needs to follow, so it is still a more straightforward setup for the agent to distinguish the sound it needs to follow. While these approaches are suitable as a starting point, it is hard to go far with them in a real-world environment.
Inspired by challenges of real-world scenarios, we design complex audio scenarios in which the agent is confronted with second sound-emitting sounds at the goal location, noisy audio sensors, and distractor sounds at different locations. This provides a more realistic, highly diverse training experience to ensure it has to focus on the directional and spatial information in the audio signal to improve generalization to unheard and noisy environments at test time.

We design highly randomized audio scenarios with perturbations and augmentations on both episode and step levels. The full randomization pipeline is shown in \cref{alg:algorithm1} and consists of three elements:
\begin{itemize}[noitemsep, topsep=0pt]
    \item Distractor Sounds: With a fixed probability, each episode includes a distractor. For each step, we randomize the distractor location and whether the distractor is audible. The distractor sound is drawn uniformly random from all training sounds, excluding the current target sound.
    \item Second Sound: For each episode, with a certain probability we overlay an additional audio signal coming from the same position over the target sound.
    \item Spectrogram Augmentations: In each step, a random augmentation is applied to the spectrogram. We draw on \cite{park2019specaugment} to construct a set of augmentations consisting of (none, time masking, frequency masking, both).
\end{itemize}

These scenarios increase the difficulty of the task, requiring the agent to reason about the elements in the audio signal. \myworries{To arrive at the correct target the agent has to learn to separate the continuously emitted target sound from the temporary and changing distractor sounds.} At the same time, \myworries{these audio scenarios} largely increase the diversity of training experience, which has shown to be very beneficial, particularly in scenarios with limited data, such as the comparatively small audio dataset of 102 sounds used here. All augmentations are purely based on the training sounds, avoiding any leaks from the validation or test sounds.

% We propose a training randomization pipeline that does the randomization in two levels: 1) episode level 2) step level.
% Starting with episode level randomizations, the pipeline will decide to
% i) whether to include a second sound at the goal location or not and if it chooses to include it, it will need to select that different audio from available training sounds randomly.
% ii) it has also to decide whether to include a distractor sound in the current episode or not. While in the step level randomization, the pipeline starts to take random decisions like 1) Whether to include a distractor sound in the current step or not, and if yes, 
% i) the pipeline needs to select different audio to act as a distractor sound randomly, then ii) it has to choose where to locate it inside the environment randomly.
% 2) The pipeline also has to decide if it is going to apply a feature-wise augmentation or not, and if yes, it has to select which type of augmentation it will use randomly; it can select whether to use (time masking, frequency masking, or a combination of both) adapted from \cite{park2019specaugment}. The full pipeline is shown in \cref{alg:algorithm1}.

\subsection{Spatial Audio-Visual Fusion}
\label{sec:fusion}

Existing work focuses on direct end-to-end reinforcement learning from sound and visual inputs to actions. AV-Nav \cite{chen2020soundspaces} individually encodes RGB-D and audio, while AV-WaN \cite{chen2020learning} further structures the inputs into a geometric and acoustic map before encoding them individually. Both then concatenate the individual features and let a standard GRU cell combine them. \cite{gan2020look} estimate a local occupancy map and use the audio signal to estimate the relative goal location, then combine these inputs by providing them to a planner to produce actions. But none of these approaches provide a clear structure to learn to combine these modalities.

Binaurally perceived spectrograms from the sound source contain a large amount of information about the space and room geometry, due to how the sound propagates through the rooms and reflects off of walls and obstacles. Previous work has shown that this information can reveal room geometries \cite{purushwalkam2020audio}.
We hypothesize that learning to extract and focus on this information, and to learn to combine it with the spatial information from geometric maps is an appropriate architectural prior for audio navigation tasks. Furthermore, we hypothesize that a structure that succeeds to focus on this part of the audio is more likely to generalize to unheard sounds and to succeed in noisy and distracting audio environments.

As AV-WaN we construct and continuously update an allocentric geometric map $G_{map}$ from the depth inputs $d_t$. The map has two channels, one for explored/unexplored and one for occupied/free space. We then propose an early fusion of learned audio features and a joint audio-spatial encoder of the geometric map and encoded audio features based on convolutional layers. A novel Spatial Audio Encoder module maps the binaural spectrogram into a spatial feature space. An Audio-Visual Encoder then convolves the channel-wise concatenation of the geometric map and encoded audio features, while also reducing it in dimensionality. On top of these features, an RNN serves as a memory component. Finally, we train the agent end-to-end with Proximal Policy Optimization (PPO) \cite{schulman2017proximal}. The overall architecture is shown in \cref{fig:architecture}. The encoders' details are shown in Tab.~\ref{tab:architecture}.

\begin{table}
  \tablesize
  \centering
  \begin{tabular}{l|l}
  \toprule
  Encoder & Structure \\
  \midrule
Depth & Conv[(8, 4), (4, 2), (3, 2)], MLP[512]\\
Audio (Replica) & Conv[(8, 4), (4, 2), (3, 1)], MLP[512]\\
Audio (MP3D) & Conv[(5, 2), (3, 2), (3, 1)], MLP[512]\\
Spatial-Audio (Replica) & ConvT[(8, 3), ((1, 13), 1)]\\
Spatial-Audio (MP3D) & ConvT[((5, 2), (3, 4)), ((4, 2), (1, 2))], \\
                     & Conv([(1, 5), 1])\\
Audio-Visual & Conv[(8, 4), (4, 2), (3, 2)], MLP[512]\\
\bottomrule
  \end{tabular}
      \caption{Architecture details. Conv indicates convolutional layers and ConvT transposed convolutions. Their arguments indicate layers according to (kernal, stride). All layers are followed by a ReLU activation. MLP denotes multi-layer perceptrons.}
      \vspace{-0.3cm}
  \label{tab:architecture}
\end{table}

\begin{table*}
  %\vspace{-0.3cm}
  \tablesize
  \centering
  \begin{tabular}{=l|+c+c+c|+c+c+c|+c+c+c|+c+c+c}
    \toprule
    Model & \multicolumn{6}{c|}{Replica} & \multicolumn{6}{c}{MP3D} \\
    \cmidrule{2-13}
    & \multicolumn{3}{c|}{Multiple Heard} & \multicolumn{3}{c|}{Unheard} & \multicolumn{3}{c|}{Multiple Heard} & \multicolumn{3}{c}{Unheard} \\
    \cmidrule{2-13}
    & SPL & SR & SNA  & SPL & SR & SNA  & SPL & SR & SNA  & SPL & SR & SNA \\
    \midrule
    % AV-Nav \cite{chen2020soundspaces} & 64.5 & -- & -- & 45.4 & -- & -- & 44.8 & -- & -- & 33.8 & -- & -- \\
    % AV-Nav \cite{chen2020learning} & -- & -- & -- & 34.7 & 50.9 & 16.7 & -- & -- & -- & 25.9 & 40.1 & 12.8 \\
    AV-Nav~\cite{chen2020soundspaces} & 55.6 & 75.4 & 29.0 & 39.4 & 57.1 & 18.8 & 51.8 & 69.5 & 31.5 & 28.2 & 40.4 & 15.9 \\ 
    % AV-WaN \cite{chen2020learning} & -- & -- & -- & 34.7 & 52.8 & 27.1 & -- & -- & -- & 40.9 & 56.7 & 30.6\\
    AV-WaN~\cite{chen2020learning} &70.4   & \textbf{90.9} & 52.5 & 29.9 & 41.6 & 23.0 & 55.4 & 82.4 & 42.5 & 36.4 & 54.6 & 27.7 \\
    %SAVi 10k & 34.3 & 45.8 & 21.6 & 22.7 & 31.7 & 13.1 & 33.6 & 42.5 & 11.2 & 27.8 & 36.5 & 9.0 \\
    %\rowstyle{\color{blue}}
    SAVi~\cite{chen2021semantic} & 45.1 & 54.0 & 30.8 & 27.5 & 33.9 & 17.2 & 29.1 & 40.3 & 13.0 & 20.4 & 29.5 & 9.6 \\
    \textbf{Ours} & \textbf{72.6}& 85.1&\textbf{54.0} & \textbf{45.6} & \textbf{58.5} & \textbf{33.4} & \textbf{61.9}& \textbf{82.9}& \textbf{46.8}& \textbf{42.4}& \textbf{55.3}& \textbf{31.6}\\
    % + specto aux & 62.8 & 85.4 & 45.6 & \textbf{59.2} & \textbf{85.2} & 42.9 & 63.5& 83.3&48 &47.8 &58.4 &34.15 \\
    % + dynamic &  &  &  & & & & 49.8&73 & 34.6&35.7 &48.6 &24.8 \\
        
    %  \textbf{spec-diff} & 65.6 & 82.9 & 48.3 & 47.4 & 64.3 & 34.6 & 59.2 & 79.5 & 44.0 & \textbf{45.9} & \textbf{60.8} & \textbf{33.9} \\
    %  \textbf{spec-diff+upscal} & 63.0 & 82.9 & 45.8 & \textbf{48.5} & \textbf{69.5} & \textbf{35.1} & \textbf{63.9} &\textbf{ 83.5} & \textbf{47.6} & 43.3 & 55.5 & 31.9 \\
    \midrule
    AV-Nav + comp &\textbf{62.6} & \textbf{85.6} & 31.3 & 45.4 & 67.0 & 21.6 & 46.3 & 68.3 & 26.3 & 30.9 & 51.2 & 16.1\\
    AV-WaN + comp & 53.1 & 80.4 & \textbf{40.1} & 41.5 & 68.8 & 31.1 & 59.1 &\textbf{86.2} & 45.3 & 41.9 & 66.2 & 31.7 \\
    %SAVi 10k + comp & 39.7 & 57.7 & 22.4 & 32.8 & 47.3 & 17.9 & 33.8 & 50.2 & 23.0 & 26.6 & 41.4 & 17.4 \\
    %\rowstyle{\color{blue}}
    SAVi + comp & 43.1 & 58.6 & 25.0 & 33.6 &  45.9 & 18.1 & 48.9 & 61.4 & 35.2 & 35.9 & 47.5 & 25.4 \\
    % + complex &  &  &  &  & & &59.8 &79.7 & 44.4 & 46.4 & 66.0 & 34.3  \\ 
    \textbf{Ours + comp} & 55.7 & 77.4 & 39.4 & \textbf{54.0} & \textbf{77.8} &\textbf{37.4} & \textbf{63.8}& 83.5 & \textbf{47.0}& \textbf{49.9} & \textbf{70.6} & \textbf{35.8}  \\ 
    % \textbf{spec-diff + comp} & 55.0 & 77.9 & 38.6 & \textbf{54.8} & \textbf{79.5} & \textbf{38.3} & 65.8 & 84.2 & 48.6 & 53.2 & \textbf{71.6} & 38.9 \\
    % \textbf{spec-diff+upscal+comp} & 53.5 & 75.8 & 38.0 & 53.7 & 78.3 & 38.1 &\textbf{ 66.0 }& 84.3 & \textbf{49.8} &\textbf{ 54.4} & 71.0 & \textbf{40.1} \\

    %\midrule
    % \textbf{Ours + comp + dyn} & 46.9 & 60.8 & 34.8 & 37.4&52.7 &27.5& 63.7& 81.6 & \textbf{48.2} & 44.2 & 63.3 & 33.6 \\

    %  + complex + aux(0.005) &  &  &  &  &  & & 60.9& 80&45.9 & 54.9 &74.3 & \textbf{40.6}  \\ 
    \bottomrule
  \end{tabular}
    \caption{Results on the \textbf{static} Audio Goal task \textbf{without} complex scenarios. The heard experiments are trained on multiple sounds and evaluated on the same sounds but in unseen environments. Unheard are trained on multiple sounds and evaluated on multiple unheard sounds in unseen environments. The +comp in the models column refers to using complex scenarios in training.}
  \label{tab:static}
%   \vspace{-0.5cm}
\end{table*}

\begin{table*}
  %\vspace{-0.3cm}
  \tablesize
  \centering
  \begin{tabular}{=l|+c+c+c|+c+c+c|+c+c+c|+c+c+c}
    \toprule
    Model & \multicolumn{6}{c|}{Replica} & \multicolumn{6}{c}{MP3D} \\
    \cmidrule{2-13}
    & \multicolumn{3}{c|}{Multiple Heard} & \multicolumn{3}{c|}{Unheard} & \multicolumn{3}{c|}{Multiple Heard} & \multicolumn{3}{c}{Unheard} \\
    \cmidrule{2-13}
    & SPL & SR & SNA  & SPL & SR & SNA  & SPL & SR & SNA  & SPL & SR & SNA \\
    \midrule
    AV-Nav + comp & \textbf{53.8} & \textbf{80.1} & 26.5 & 44.4 & 67.7 & 21.7 & 41.5 & 66.3 & 22.5 & 33.3 & 55.5 & 17.6\\
    AV-WaN + comp  & 46.7 & 75.3 & 35.0 & 39.4 & 65.2 & 29.2 & 56.7 & \textbf{83.9 }& 43.9 & 46.4 & 70.8 & 35.2  \\
    % + complex &  &  &  &  & & &55.6 &76.9 & 41.5& 47.8 & 68.0 & 35.6  \\ 
    %SAVi 10k + comp & 33.9 & 51.1 & 18.8 & 27.1 & 41.7 & 15.0 & 34.4 & 53.8 & 22.5 & 30.6 & 48.9 & 19.6 \\
    %\rowstyle{\color{blue}} 
    SAVi + comp & 34.9 & 51.0 & 20.4 & 30.0 & 42.8 & 16.6 & 44.2 & 59.9 & 31.8 & 37.6 & 52.9 & 26.6 \\
    \textbf{Ours + comp} & 50.5 & 74.5 & \textbf{35.8} &  \textbf{49.9}& \textbf{74.8}& \textbf{35.4}& \textbf{60.1} & 82.2 & \textbf{44.4} & \textbf{52.6} & \textbf{72.5} & \textbf{37.9} \\
    % \textbf{spec-diff+comp} & 50.2 & 73.4 & 35.3 & \textbf{50.4} & \textbf{75.2} & \textbf{35.5} & 61.3 & 82.7 & 45.6 & 54.2 & \textbf{74.4} & 40.1 \\
    % \textbf{spec-diff+upscal+comp} & 49.6 & 72.3 & 35.4 & 48.6 & 72.1 & 34.7 & \textbf{61.8} & 81.5 & \textbf{46.5} & \textbf{54.7} & 74.3 & \textbf{40.8} \\
    \bottomrule
  \end{tabular}
    \caption{Results on the \textbf{static} Audio Goal task \textbf{with} complex scenarios. %The heard experiments are trained on multiple sounds and evaluated on the same sounds but in unseen environments. Unheard are trained on multiple sounds and evaluated on multiple unheard sounds in unseen environments.
    }
  \label{tab:static_complex}
  \vspace{-0.3cm}
\end{table*}

\textit{Action Parametrization}: While AV-Nav \cite{chen2020soundspaces} reasons directly in the raw action space of the simulator, AV-WaN \cite{chen2020learning} demonstrated further improvements from learning to select waypoints on a higher level of abstraction. The agent chooses from a $9\times9$ action map centered on the agent's current position. A simple Dijkstra planner then navigates to this waypoint. While acting in such a lifted Markov Decision Process can be beneficial, far away waypoints correspond to a lower control frequency, ignoring up to ten observations while the planner is executing the actions to reach the selected waypoint. While the loss of information from these observations might be negligible in scenarios with a clean, static sound source, it becomes much more important to continuously integrate the audio observations over time in the noisy and dynamic audio scenarios that we present in this paper. 
Correspondingly we find it beneficial to decrease the size of the action map to $3\times3$, providing an efficient middle ground between the benefits of learning waypoints and decreasing the number of unprocessed observations to a maximum of four.

\section{Experimental Evaluation}\label{sec:experiments}
% \input{tables/experiments}
% \input{tables/complex_scenarios}

% \comm{Levels of task difficulty to construct tasks from (and combinations thereof) for 2 datasets: replica and MP3D:
% \begin{itemize}
%     \item Heard sounds
%     \item Unheard sounds
%     \item Unseen apartments
%     \item With distractors
%     \item Moving
% \end{itemize}
% } 

%In this section, we evaluate both existing approaches and our proposed architecture on the static and dynamic AudioGoal tasks and study the impact of training and evaluating in complex audio scenarios.

\subsection{Task Setups}

We tackle the tasks of static and dynamic AudioGoal navigation. For each task, we train all agents in two scenarios: the clean audio setup used in \cite{chen2020soundspaces,chen2020learning} and the complex audio scenarios with noisy audio, distractors, and second sound-emitting source that we introduced in \cref{sec:complex}. All agents are trained on multiple sounds and evaluated in two settings: on heard sounds in unseen environments and on unheard sounds in unseen environments. We use the same train/val/test splits protocol used by \cite{chen2020soundspaces,chen2021learning}, where Replica splits into 9/4/5 scenes and Matterport3D (MP3D) splits into 59/10/12 scenes. The 102 different sounds are split into 73/11/18. The same split is applied to any other audio signals, such as distractors. This signifies that evaluations on unheard sounds are also required to handle unheard disturbances. We report the mean over three models trained with different random seeds.

\begin{table*}
  %\vspace{-0.3cm}
  \tablesize
  \centering
  \begin{tabular}{=l|+c+c+c|+c+c+c|+c+c+c|+c+c+c}
    \toprule
    Model & \multicolumn{6}{c|}{Replica} & \multicolumn{6}{c}{MP3D} \\
    \cmidrule{2-13}
    & \multicolumn{3}{c|}{Multiple Heard} & \multicolumn{3}{c|}{Unheard} & \multicolumn{3}{c|}{Multiple Heard} & \multicolumn{3}{c}{Unheard} \\
    \cmidrule{2-13}
    & DSPL & SR & DSNA  & DSPL & SR & DSNA  & DSPL & SR & DSNA  & DSPL & SR & DSNA \\
    \midrule
    AV-Nav~\cite{chen2020soundspaces} & 47.2 &  71.0 & 16.3 & 23.4 & 37.4 & ~7.8  & 43.4 & 73.0 & 18.4 & 20.9 & 35.5 & ~8.4 \\ 
    AV-WaN~\cite{chen2020learning} &  55.0 & 86.8 & 32.6 & 24.1 & 38.2 & 14.2 & 58.3 & 86.7 & 33.6 &  30.0 & 47.2 & 17.4 \\
    %\rowstyle{\color{blue}}
    SAVi~\cite{chen2021semantic} & 54.0 & 67.6 & 25.7 & 28.9 & 40.8 & 13.9 & 29.9 & 49.0 & 11.3 & 26.3 & 41.6 & 10.0 \\
    % \rowstyle{\color{blue}}ours +  new R & 61.0 & 87.2 & 33.9 & 32.8 & 48.9 & 18.1 & 49.6 & 89.1 & 30.9 & 30.4 & 52.1 & 19.1 \\
    \textbf{Ours} & \textbf{62.6} & \textbf{90.0} & \textbf{34.5} & \textbf{29.4} & \textbf{49.1} & \textbf{16.1} & \textbf{63.2} & \textbf{94.0} & \textbf{35.3} & \textbf{37.0}   & \textbf{57.7} & \textbf{21.0}\\
    % +aux & 58 & 89.4 &  &  28&  45.5&  & 51.9 &88.3  &  &  38.9 & 63.4 &\\
    % \textbf{spec-diff} & 59.4 & 83.5 & 33.3 & 32.5 & 51.0 & 17.9 & 59.6 & 92.0 & 34.9 & 39.0 & 60.5 & 22.4  \\
    % \textbf{spec-diff+upscal} & 61.0 & 82.9 & 33.9 & \textbf{33.6} & \textbf{52.1} & \textbf{18.4} & 63.0 & \textbf{94.1} & \textbf{36.5} & \textbf{41.9} &\textbf{ 63.8} & \textbf{24.2}  \\
    \midrule
    % + complex scenarios &  &  &  & 25.1 & 55.2 &  &  &  &  &  &  &   \\ 
    AV-Nav + comp& 42.1 & 66.4 & 14.5 & 25.6 & 42.5 & ~8.5 & 37.6 & 66.5 & 15.6 & 21.2 & 37.8 & ~8.3 \\
    AV-WaN + comp & \textbf{46.4} & \textbf{75.3} & \textbf{27.7} & 27.9 & 53.2 & 16.6 & 56.0 & 87.0 & 33.7 & 35.2 & 58.4 & 21.2\\
    %\rowstyle{\color{blue}}
    SAVi + comp & 45.0 & 57.9 & 20.9 & 29.2 & 41.7 & 13.1 & 47.5 & 67.4 & 24.0 & 29.9 & 45.5 & 15.1 \\
    %\rowstyle{\color{blue}}ours + comp + new R & 42.3 & 67.8 & 23.5 & 31.0 & 58.1 & 17.2 & 63.5 & 94.7 & 38.0 & 41.2 & 68.4 & 24.7 \\
    \textbf{Ours + comp} & 43.9 & 67.6 & 24.5 & \textbf{33.9} & \textbf{67.3} & \textbf{19.1 }&  \textbf{61.6} &  \textbf{91.9} & \textbf{36.2} & \textbf{45.8} & \textbf{74.0} & \textbf{26.0}  \\ 

    % \textbf{spec-diff+comp} & 37.1 & 63.9 & 20.8 & 28.5 & 57.0 & 15.8 & 61.0 & \textbf{93.1} & 35.3 & \textbf{47.4} & \textbf{78.0} & \textbf{27.6}  \\
    % \textbf{spec-diff+upscal+comp} & 42.6 & 66.7 & 23.8 & \textbf{34.4}& 65.2 & \textbf{19.1} & 60.8 & 91.3 & 35.1 & 47.3 & \textbf{78.0} & 27.2 \\
    \bottomrule
  \end{tabular}
%   \vspace{-0.5cm}
  \caption{Results on the \textbf{dynamic} Audio Goal task \textbf{without} complex scenarios. %The heard experiments are trained on multiple sounds and evaluated on the same sounds but in unseen environments. Unheard are trained on multiple sounds and evaluated on multiple unheard sounds in unseen environments.
  }
  \label{tab:dynamic}
  \vspace{-0.3cm}
\end{table*}

\noindent\textit{Metrics}: We evaluate the navigation performance based on:
\begin{itemize}[noitemsep, topsep=0pt]
\item \textit{Success rate (SR)}: The share of successful episodes of all test episodes. An episode is considered successful if the agent executes the $stop$ action at the goal location.
\item \textit{Success weighted by path length (SPL) \cite{anderson2018evaluation}}: The ratio of the length of the shortest path to the goal to the length of the executed path for the successful episodes.
\item \textit{Success weighted by number of actions (SNA) \cite{chen2020learning}}: The ratio of the number of actions needed to follow the shortest path to the actual number of actions the agents took to reach the same goal. In contrast to the SPL this metric takes the number of orientation changes into account.\looseness=-1
\item \textit{Dynamic success weighted by path length (DSPL)}: The primary metric for dynamic AudioGoal Navigation. It is calculated as the ratio of the length of the path to the earliest reachable intersection and the length of the executed path for the successful episodes, see \cref{sec:dynamic_task}.
\item \textit{Dynamic success weighted by number of actions (DSNA)}:
Equivalently to the DSPL, we calculate an adjusted version of the SNA with respect to the same definition of the earliest reachable intersection.
% We use an updated version of the SNA for the Dynamic Audio-Visual Navigation task, as we do not have prior knowledge about the least amount of actions needed to reach the goal. Therefore, we rely on the DSPL to get the closest intersection point; then, we calculate the number of actions the agent needs to reach this point. Finally, we calculate it as the ratio between that number of actions and the actual number of actions the agent executed during successful episodes.
\end{itemize}

{\parskip=5pt
\noindent\textit{Baselines}: We compare our approach to \myworries{three} current state-of-the-art methods: \textit{AV-Nav}~\cite{chen2020soundspaces} is an end-to-end reinforcement learning agent that directly encodes audio and visual observations to select actions using audio-visual observations. \textit{AV-WaN}~\cite{chen2020learning} is the current state-of-the-art for static AudioGoal task, which predicts intermediate waypoints to the goal depending on audio observation, geometric, and acoustic maps. For both models, we use the author's code. \myworries{\textit{SAVi}~\cite{chen2021semantic} is a transformer based model originally developed for semantic audio-visual navigation. As our task does not have a semantic aspect and sound is emitted throughout the whole episode, we train it without the goal descriptor network.} Hyperparameters for all trained models are included in the supplementary material. To decompose our contributions and ensure a fair comparison, we evaluate all models with and without training on complex audio scenarios. Rows marked "+comp" report training in complex scenarios.}

\subsection{Static AudioGoal Task}

\begin{table*}
%   \vspace{-0.3cm}
  \tablesize
  \centering
  \begin{tabular}{=l|+c+c+c|+c+c+c|+c+c+c|+c+c+c}
    \toprule
    Model & \multicolumn{6}{c|}{Replica} & \multicolumn{6}{c}{MP3D} \\
    \cmidrule{2-13}
    & \multicolumn{3}{c|}{Multiple Heard} & \multicolumn{3}{c|}{Unheard} & \multicolumn{3}{c|}{Multiple Heard} & \multicolumn{3}{c}{Unheard} \\
    \cmidrule{2-13}
    & DSPL & SR & DSNA  & DSPL & SR & DSNA  & DSPL & SR & DSNA  & DSPL & SR & DSNA \\
    \midrule
    AV-Nav + comp & 34.0 & 60.4 & 11.5 & 26.0 & 45.4 & ~8.7 & 34.4 & 63.2 & 13.4 & 24.6 & 46.2 & ~9.3\\ 
    AV-WaN + comp & \textbf{43.1} & \textbf{66.3} & \textbf{25.1} & 32.8 & 51.5 & \textbf{19.0} & 52.3 & 84.0 & 31.5 & 43.0 & 70.3 & 25.6\\
    %\rowstyle{\color{blue}}
    SAVi + comp & 34.2 & 45.5 & 16.1 & 27.7 & 38.3 & 12.7 & 44.0 & 65.6 & 22.0 & 35.1 & 55.5 & 16.0 \\
    %\rowstyle{\color{blue}}ours + comp + new R & 33.1 & 55.8 & 18.4 & 29.3 & 49.2 & 16.2 & 54.9 & 93.3 & 32.6 & 44.8 & 77.0 & 27.0 \\
    \textbf{Ours + comp} & 38.6 & 63.8 & 21.8 & \textbf{32.9} &\textbf{58.9}  & 18.6  & \textbf{59.9} & \textbf{92.8} & \textbf{33.3} & \textbf{52.1} & \textbf{84.1}& \textbf{30.4}  \\ 
    % \textbf{spec+comp} & 31.3 & 56.4 & 17.4 & 26.6 & 51.9 & 14.5  & 57.4 & \textbf{93.1} & \textbf{33.5} & 51.0 & 83.9 & 29.5  \\ 
    % \textbf{spec+upscal+comp} & 37.2 & 61.6 & 20.7 & \textbf{33.6} & 58.0 & 18.6  & 57.3 & 92.3 & 33.0 & 50.4 & \textbf{85.1} & 29.3\\ 
    \bottomrule
  \end{tabular}
%   \vspace{-0.5cm}
  \caption{Results on the \textbf{dynamic} Audio Goal task \textbf{with} complex scenarios. %The heard experiments are trained on multiple sounds and evaluated on the same sounds but in unseen environments. Unheard are trained on multiple sounds and evaluated on multiple unheard sounds in unseen environments.
  }
  \label{tab:dynamic_complex}
\end{table*}

\begin{figure*}
    % \vspace{-0.3cm}
    \centering
    \footnotesize
    \setlength{\tabcolsep}{0.0cm}% for the horiz padding
    {\renewcommand{\arraystretch}{0.1}% for the vertical padding
    \newcolumntype{M}[1]{>{\centering\arraybackslash}m{#1}}
    \begin{tabular}{M{3.0cm}M{3.0cm}M{3.8cm}M{3.8cm}}
    \multicolumn{2}{c}{Replica} & \multicolumn{2}{c}{Matterport3D} \\
    \includegraphics[width=0.95\linewidth,trim={2.2cm 1.5cm 2.2cm 1.7cm},clip]{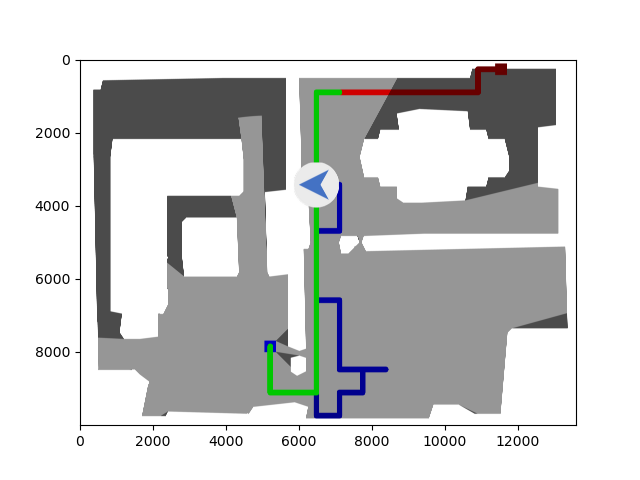} & \includegraphics[width=0.95\linewidth,trim={2.2cm 1.5cm 2.2cm 1.7cm},clip]{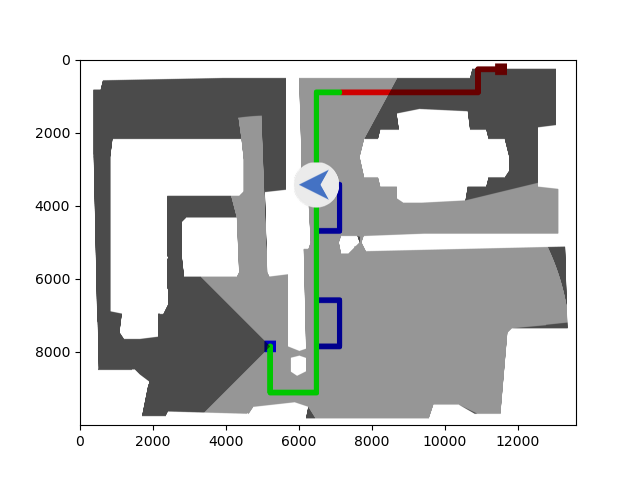} & \includegraphics[width=1\linewidth,trim={2.2cm 1.4cm 2.2cm 3.4cm},clip]{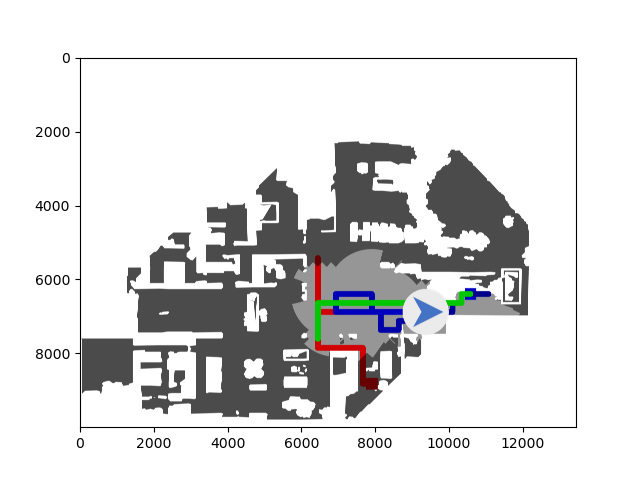} & \includegraphics[width=1\linewidth,trim={2.2cm 1.4cm 2.2cm 3.4cm},clip]{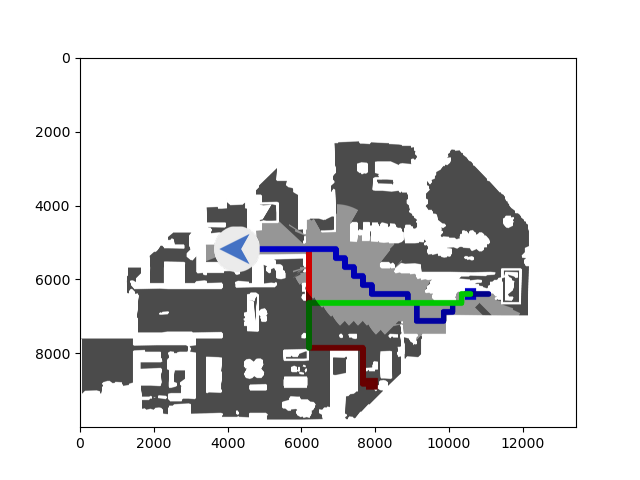} \\
    \\
    \\
    {(a) AV-WaN \cite{chen2020learning}} & {(b) Ours} & {(c) AV-WaN \cite{chen2020learning}} & {(d) Ours}
    \end{tabular}}
    % \vspace{-0.2cm}
    \caption{Example episodes of the heard dynamic audio-visual navigation task on Replica (left) and Matterport3D (right). For each, the AV-WaN agent is shown on the left, and our architecture on the right, trained without complex scenarios. The paths of the agent and sound source are shown in blue and red, respectively. Green shows the path to the earliest reachable intersection as defined for the DSPL metric.}
    \label{fig:example}
    %\vspace{-0.3cm}
\end{figure*}

We first evaluate the static AudioGoal task, in which the agents are trained on multiple heard sounds and evaluated in unseen apartments with either the heard or unheard sounds. The upper half of \cref{tab:static} shows the results for the original setup without complex audio scenarios. Our proposed novel channel for a learned spatial fusion achieves similar or even better performance on the heard sounds and significantly improves on unheard sounds. While AV-Nav and AV-WaN each perform well on one of the datasets, they underperform on the other with differences of up to 17ppt in success rates. Our model achieves the consistently best performance across both datasets. This transfers to an even larger difference in path optimality as measured by SPL and SNA.
The bottom part of the table shows the effects of training on the complex audio scenarios described in \cref{sec:complex} and evaluation on the clean, standard benchmark ("+ comp" rows). Training in these scenarios provides extensive improvements for the generalization to unheard sounds across all models, with improvements of over 19ppt for some models. Again, our model achieves the highest performance across all metrics. Combined, our new architecture and the complex scenarios increase the performance from a success rate of 57.1 to 77.8 on Replica and 54.6 to 70.6 on MP3D, strongly outperforming previous state-of-the-art results on this benchmark.
A further decomposition of the effects of the individual elements of the complex scenarios can be found in the supplemental material.
Performance of our architecture decreases when trained on complex scenarios and evaluated only on the heard sounds in Replica. This can be attributed to the small dataset size or the complex scenarios remove certain channels that the model could previously exploit to overfit on heard sounds. 
% We then further evaluate the impact of including dynamic target sounds within the training ("+ dyn"). We let the sound source move in a random selection of half of the episodes and remain static in the other half to test the model's ability to learn both tasks simultaneously. While we observe a negative impact of this setup on Replica, performance on MP3D remains stable. The differences between moving and static sounds can be more considerable on the much smaller apartments in Replica. 
We proceed to evaluate the performance on the complex audio scenarios. Results are shown in \cref{tab:static_complex}. We find that the impact of these scenarios is larger in Replica than on the large MP3D dataset. Again we find that our architecture consistently generalizes best to unheard sounds.

\subsection{Dynamic AudioGoal Task}

We then evaluate the models on the moving sound task. All models are trained on multiple heard sounds with half of the episodes containing a moving and the other half a static sound source. The models are then evaluated on the dynamic sounds for both heard and unheard sounds. Results are shown in \cref{tab:dynamic}. Overall, we find that all models are able to solve a significant share of the tasks, with success rates of 67-94\% on heard sounds. This is roughly similar to the performance on static sound. But we find a much larger gap in the performance on unheard sounds, both in terms of success rates and optimality of the paths (SPL vs DSPL).
In terms of overall performance, we find that our architecture again clearly performs best on all clean audio setups, improving success rates by over 10ppt on unheard, dynamic sounds on both datasets.
Training on the complex scenarios again proves very beneficial for generalization to unheard sounds for all models except AV-Nav. The combination of our architecture and the complex scenarios achieves large improvements from a success rate of 38.2 for the best baseline on clean scenarios to 67.3 on replica and from 47.2 to 74.0 on MP3D. \cref{tab:dynamic_complex} evaluates the same models on the complex audio scenarios. Again, our approach consistently generalizes best with the single exception of DSNA metric on Replica. We furthermore found that removing the audio-spatial encoder leads to a clear drop in performance in the dynamic audio goal task, while the impact of removing either of the other two encoders was more limited (see Section 5 in the supplementary material).\looseness=-1
%\myworries{(see \cref{sec:model_components_supp})}.\looseness=-1
% \include{tables/architecture}

\cref{fig:example} depicts example episodes for AV-WaN and our agent on heard sounds. Optimal behavior can significantly differ from simply moving directly toward the initial position of the sound. Furthermore, acting suboptimally early on can have a large impact on the required path later on and agents may have to quickly change direction if the target moves past them. Additional examples can be found in the supplementary material.\looseness=-1
%Visualisations of the qualitative behavior of the agents on the dynamic task can be found in the supplementary material.

% \todo{Reference to augmentation\_decomposition table in the supplement}
%\input{tables/augmentation_decomposition}

\section{Conclusion}
We introduce the novel dynamic audio-visual navigation benchmark together with a novel metric that quantifies the gap to optimal behavior. We demonstrate that this task poses new challenges over existing benchmarks. We then introduce complex audio scenarios based on audio-specific augmentations, perturbations, and randomizations and demonstrate that this provides substantial benefits in generalization to unheard sounds. Lastly, we introduce an architecture with an inductive bias to spatially fuse the geometric information inherent in the audio and visual observations and demonstrate that this consistently outperforms previous approaches. Combined, this results in very large overall improvements for the generalization to unheard sounds. %while dropping most of the assumptions used by prior work.. %Our contributions improves the state-of-the-art models while dropping most of the assumptions used by prior work.

\footnotesize
\bibliographystyle{IEEEtran}
\bibliography{egbib}

\clearpage
\renewcommand{\baselinestretch}{1}
\setlength{\belowcaptionskip}{0pt}

\begin{strip}
\begin{center}
\vspace{-5ex}
\textbf{\LARGE \bf
Catch Me If You Hear Me: Audio-Visual Navigation in Complex\\Unmapped Environments with Moving Sounds} \\
\vspace{2ex}

\Large{\bf- Supplementary Material -}\\
\vspace{0.4cm}
\normalsize{Abdelrahman Younes$^*$\hspace{1cm} Daniel Honerkamp$^*$ \hspace{1cm} Tim Welschehold\hspace{1cm} Abhinav Valada}\\
\end{center}
\end{strip}

%%%%%%%%%% Merge with supplemental materials %%%%%%%%%%
%%%%%%%%%% Prefix a "S" to all equations, figures, tables and reset the counter %%%%%%%%%%
\markboth{}{}
\setcounter{section}{0}
\setcounter{equation}{0}
\setcounter{figure}{0}
\setcounter{table}{0}
\makeatletter

\renewcommand{\thesection}{S.\arabic{section}}
\renewcommand{\thesubsection}{S.\arabic{subsection}}
\renewcommand{\thetable}{S.\arabic{table}}
\renewcommand{\thefigure}{S.\arabic{figure}}
\normalsize
\begin{strip}
    \centering
    \captionsetup{type=figure}
    \setlength{\tabcolsep}{0.25cm}% for the horiz padding
    {\renewcommand{\arraystretch}{0.5}% for the vertical padding
    \newcolumntype{M}[1]{>{\centering\arraybackslash}m{#1}}
    \begin{tabular}{M{4.5cm}M{4.5cm}M{4.5cm}}
    \includegraphics[width=1\linewidth]{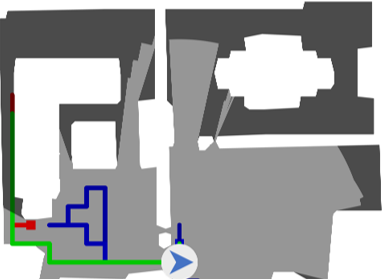} & \includegraphics[width=1\linewidth]{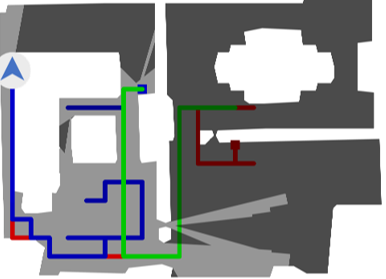} &
    \includegraphics[width=1\linewidth]{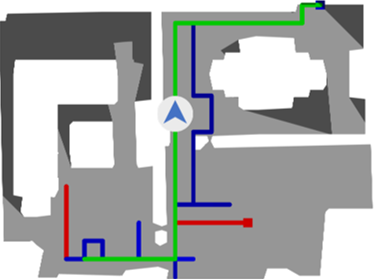} \\
    \multicolumn{3}{c}{(a) AV-WaN~\cite{chen2020learning}} \\
    \\
    \includegraphics[width=1\linewidth]{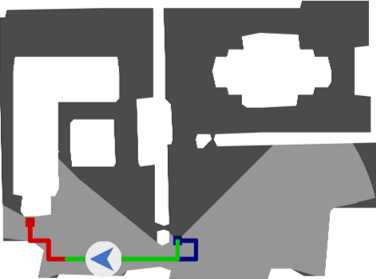} & 
    \includegraphics[width=1\linewidth]{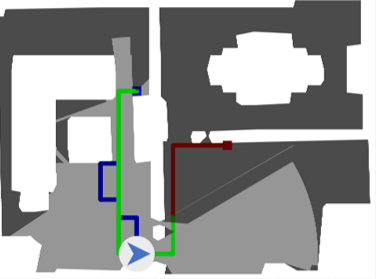} &
    \includegraphics[width=1\linewidth]{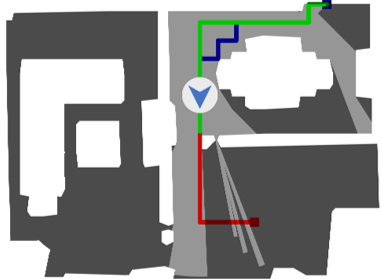} \\
    \multicolumn{3}{c}{(b) Ours}\\
    \end{tabular}}
    \captionof{figure}{Qualitative comparison of the dynamic audio-visual navigation task on the Replica dataset for heard sounds without complex scenarios. Each column shows the same episode for the AV-WaN agent on top and ours at the bottom. The paths of the agent and sound source are shown in blue and red, respectively. Start poses are marked with rectangles. Green shows the path to the earliest reachable intersection as defined for the DSPL metric.}
    \label{fig:comparison}
\end{strip}%

In this supplementary material, we analyze the impact of individual components of the complex audio scenarios and the movement speed of the dynamic sound sources in ablation studies. Moreover, we provide additional details on the hyperparameters for all models. \looseness=-1
% \begin{enumerate}[noitemsep]
%     \item Code to run the experiments for our model and all baselines.
%     \item Video for qualitative assessment of the performance of our agent against the other baselines\cite{chen2020soundspaces,chen2020learning}.
%     \item Model architecture details
%     \item The effect of the individual components of the complex scenarios. 
%     \item Hyperparameters used for training
% \end{enumerate}

% \section{Code}

% The code to run our experiments, including all reported baselines, is included as a zip file and will be made publicly available after acceptance. Please follow the instruction in the \emph{readme.md} file to reproduce our results.

% \section{Video}

% In this video, we demonstrate the novel task of Dynamic Audio-Visual Navigation and show the qualitative behavior of the agents. Please use headphones to listen to the binaural audio that the agent receives.

\section{Qualitative Results}

\cref{fig:comparison} depicts further qualitative examples for the AV-WaN~\cite{chen2020learning} agent and our model on the dynamic audio-visual navigation benchmark. Each column shows the same episode for each agent. While both agents initially move in the right direction, the AV-WaN agent ultimately has to move through a much larger part of the apartment, chasing the sound source. In contrast, our approach captures the target in its path. Novel challenges of this task include the need to adapt directions to the moving sound and potentially much higher costs for missing the sound source, which then may move further away from the agent.

\begin{table*}
%  \tablesize
  \centering
  \begin{tabular}{l|ccc|ccc|ccc|ccc}
    \toprule
    Model & \multicolumn{6}{c|}{Replica} & \multicolumn{6}{c}{MP3D} \\
    \cmidrule{2-13}
    & \multicolumn{3}{c|}{Multiple Heard} & \multicolumn{3}{c|}{Unheard} & \multicolumn{3}{c|}{Multiple Heard} & \multicolumn{3}{c}{Unheard} \\
    \cmidrule{2-13}
    & SPL & SR & SNA  & SPL & SR & SNA  & SPL & SR & SNA  & SPL & SR & SNA \\
    \midrule
    Ours & \textbf{71.9}& \textbf{85.9}&\textbf{53.7} & 48.6 & 63.6 & 35.4 & 66.2& \textbf{86.7}& 48.5& 46.3& 60.6& 33.8\\
    Ours + 2nd audio & 70.2 & 85.3 & 51.6 & \textbf{56.0} & 74.6 & \textbf{40.4} & 62.6 & 81.8 & 47.4 & 51.2 & 65.0 & 38.2 \\
    Ours + specaug & 56.5 & 78.4 & 40.2 & 53.9 & 78.7 & 38.1 & 59.2 & 79.7 & 43.3 & 39.6 & 58.3 & 28.6 \\
    Ours + 2nd audio + specaug & 59.2 & 82.4 & 41.8 & 54.5 & \textbf{80.0} & 38.3 & 62.6 & 83.2 & \textbf{48.7} & 53.0 & \textbf{73.3} & \textbf{40.7} \\ 
    Ours + 2nd audio + distractor & 61.5& 82.2 & 44.1& 48.7 & 72.1 & 34.6 & 64.7 & 83.3 & 47.8 & 52.0 & 70.1 & 37.8\\
    Ours + complex & 51.2 & 75.6 & 35.2 & 54.0 & 78.0 &36.7 & \textbf{66.3}& 86.4& 47.7& \textbf{55.0} & 73.0 & 38.6  \\ 
    \bottomrule
  \end{tabular}
      \caption{Decomposition of complex scenarios: Evaluation on the \textbf{static} Audio Goal task \textbf{without} complex scenarios.}
  \label{tab:decomp}
\end{table*}

% \begin{table}
%   \footnotesize
%   \centering
%   \begin{tabular}{@{}l|ccc|ccc@{}}
%     \toprule
%   Model & \multicolumn{6}{c}{Replica} \\
%     & & Heard & & & Unheard \\
%      & SPL & SR & SNA & SPL & SR & SNA   \\
%     \midrule
%     Default & 62.8 & 85.4 & 45.6 & \textbf{59.2} & \textbf{85.2} & 42.9 \\
%     + augment & &  &  & 56.1 & 82.7 & 40 \\
%     + secondAudio & &  & &  58.2 & 82.7 & 41.9  \\
%     + specaugment + secondAudio & &  & &  57.4 & 85.2 & 40.9  \\
%     + distractor & \\
%     + train with moving & \\

%     \bottomrule
%   \end{tabular}
%   \caption{}
%   \label{tab:complex}
% \end{table}

\begin{table*}
%  \tablesize
  \centering
  \begin{tabular}{l|ccc|ccc|ccc|ccc}
    \toprule
    Model & \multicolumn{6}{c|}{Replica} & \multicolumn{6}{c}{MP3D} \\
    \cmidrule{2-13}
    & \multicolumn{3}{c|}{Multiple Heard} & \multicolumn{3}{c|}{Unheard} & \multicolumn{3}{c|}{Multiple Heard} & \multicolumn{3}{c}{Unheard} \\
    \cmidrule{2-13}
    & SPL & SR & SNA  & SPL & SR & SNA  & SPL & SR & SNA  & SPL & SR & SNA \\
    \midrule
    Ours& 24.5&31.7&18.0&20.0&26.9&15.0&40.0&51.3&28.7&32.1&43.1&23.4\\
    Ours + 2nd audio & 31.0 & 41.7 & 22.6 & 23.6 & 33.1 & 16.8 & 46.9 & 62.1 & 35.8 & 43.1 & 56.5 & 32.8 \\
    Ours + specaug & 47.1 & 67.7 & 33.3 & 45.7 & 67.8 & 32.3 & 50.7 & 70.9 & 36.5 & 40.4  & 58.0 & 28.7 \\
    Ours + 2nd audio + specaug & \textbf{49.3} & 70.4 & \textbf{34.9} & 46.1 & 68.8 & 32.6 & 55.6 & 79.0 & 43.3 & 53.4 & 72.9 & \textbf{40.7} \\ 
    Ours + 2nd audio + distractor & 38.5& 56.1 & 27.6& 32.9 & 47.9 & 23.4 & 49.1 & 64.9 & 36.8 & 41.5 & 55.3 & 30.6\\
    Ours + complex & 47.9 & \textbf{73.7} & 33.1 &  \textbf{47.8} & \textbf{73.4} & \textbf{33.3} & \textbf{64.1} & \textbf{86.4} & \textbf{46.8}& \textbf{55.5} & \textbf{76.0} & 40.4 \\
    \bottomrule
  \end{tabular}
    \caption{Decomposition of complex scenarios: Evaluation on the \textbf{static} Audio Goal task \textbf{with} complex scenarios.}
  \label{tab:decomp_complex}
\end{table*}
\begin{table*}
  \tablesize
  \centering
  \begin{tabular}{l|ccc|ccc|ccc|ccc}
    \toprule
    Model & \multicolumn{6}{c|}{Replica} & \multicolumn{6}{c}{MP3D} \\
    \cmidrule{2-13}
    & \multicolumn{3}{c|}{Multiple Heard} & \multicolumn{3}{c|}{Unheard} & \multicolumn{3}{c|}{Multiple Heard} & \multicolumn{3}{c}{Unheard} \\
    \cmidrule{2-13}
    & DSPL & SR & DSNA  & DSPL & SR & DSNA  & DSPL & SR & DSNA  & DSPL & SR & DSNA \\
    \midrule
    Ours $p=10\%$ & \textbf{68.3} & \textbf{88.9} & \textbf{36.4} & \textbf{40.2} & \textbf{67.3} & \textbf{21.4} & \textbf{69.3} & 93.6 & \textbf{38.4 } & \textbf{49.0} & 68.9 & 27.1\\
    Ours $p=20\%$  & 66.2 & 87.3 & 36.0 & 38.2& 64.2&20.3 & 67.8 & \textbf{93.7} & 37.9 & 48.2 & \textbf{69.7} & \textbf{27.2}  \\
    % + complex &  &  &  &  & & &55.6 &76.9 & 41.5& 47.8 & 68.0 & 35.6  \\ 
     Ours $p=30\%$  & 58.0 & 85.1 & 32.3 & 33.7 & 58.7 & 18.5 & 61.2 & 92.0 & 34.7 & 43.4 & 68.1 & 24.8  \\
      Ours $p=40\%$  & 44.3 & 80.5 & 25.1 & 23.7 & 49.8 & 13.5 & 46.5 & 89.8 & 26.6 & 31.3 & 64 & 18.1 \\
    \bottomrule
  \end{tabular}
    \caption{Motion model ablation: The percentages represent the sound source moving percentage during test time. Trained on static + dynamic with $p=30\%$. Results on the \textbf{dynamic} Audio Goal task \textbf{without} complex scenarios.}
  \label{tab:moving_percentage}
\end{table*}

% \begin{table}
%   \tablesize
%   \centering
%   \begin{tabular}{l|c}
%   \toprule
% Hyperparameter&Value \\\midrule
% clip param & 0.1 \\
% ppo epoch & 4 \\
% num mini batch & 1 \\
% value loss coef & 0.5 \\
% entropy coef (AV-Nav) & 0.02 (0.2) \\
% lr & $\expnumber{2.5}{-4}$ \\
% eps & $\expnumber{1}{-5}$ \\
% max grad norm & 0.5 \\
% optimizer & Adam \\

% steps number & 150 \\
% gru hidden size& 512 \\
% use gae& True \\
% gamma& 0.99 \\
% tau& 0.95 \\
% linear clip decay&True \\
% linear lr decay& True \\
% exponential lr decay& False \\
% exp decay lambda& 5.0 \\
% reward window size& 50 \\
% \midrule
% number of processes (AV-Nav Matterport3d) & 5 (10) \\
% number of updates (AV-Nav) & 10,000 (40,000)\\
% \midrule
% time mask param (Replica) & 32\\
% frequency mask param (Replica) & 12\\
% time mask param (Matterport3d) & 12\\
% frequency mask param (Matterport3d) &12\\
% \bottomrule
%   \end{tabular}
%       \caption{Hyperparameters used for training our model and the baselines. Differences across models are shown in parentheses.}
%   \label{tab:hyper}
% \end{table}

\begin{table*}
  \tablesize
  \centering
  \begin{tabular}{l|c||l|c}
  \toprule
Hyperparameter&Value&Hyperparameter&Value \\
\midrule
clip param & 0.1 & optimizer & Adam \\
ppo epoch & 4 & steps number & 150 \\
num mini batch & 1 & gru hidden size& 512 \\
value loss coef & 0.5 & use gae& True \\
entropy coef (AV-Nav) & 0.02 (0.2) & gamma& 0.99 \\
lr & $\expnumber{2.5}{-4}$ & tau& 0.95 \\
eps & $\expnumber{1}{-5}$ & linear clip decay&True \\
linear lr decay& True & exponential lr decay& False \\
exp decay lambda& 5.0 & max grad norm & 0.5 \\
reward window size& 50 & & \\
\midrule
number of processes (AV-Nav MP3D) & 5 (10) & number of updates (AV-Nav) & 10,000 (40,000)\\
\midrule
time mask param (Replica) & 32 & frequency mask param (Replica) & 12\\
time mask param (MP3D) & 12 & frequency mask param (MP3D) &12\\
\bottomrule
  \end{tabular}
      \caption{Hyperparameters used for training our model and the baselines. Differences across models are shown in parentheses. SAVi uses the parameters reported in~\cite{chen2021semantic}.}
  \label{tab:hyper}
\end{table*}
% \begin{table*}
%   \vspace{-0.3cm}
%  % \tablesize
%   \centering
%   \begin{tabular}{l|ccc|ccc|ccc|ccc}
%     \toprule
%     Model & \multicolumn{6}{c|}{without complex} & \multicolumn{6}{c}{with complex} \\
%     \cmidrule{2-13}
%     & \multicolumn{3}{c|}{Multiple Heard} & \multicolumn{3}{c|}{Unheard} & \multicolumn{3}{c|}{Multiple Heard} & \multicolumn{3}{c}{Unheard}\\
%     \cmidrule{2-13}
%     & SPL & SR & SNA  & SPL & SR & SNA  & SPL & SR & SNA  & SPL & SR & SNA \\
%     \midrule
%      % w/o Depth &\textbf{ 64.9 } & \textbf{83.8  } & \textbf{47.8 } & \textbf{53.6 } & \textbf{72.9} & \textbf{38.8} & \textbf{60.1 } & 80.9  & 44.1 & \textbf{54.5 } & \textbf{ 74.7 } & \textbf{39.5 } & 1.2\\
%      w/o Audio  & 57.9  & 78.2 & 42.9 & 49.7 & \textbf{71.6} & \textbf{36.7} & 55.6 & 78.8 & 41.5 & 49.6 & 72.1 & 37.0 \\
%      w/o  Audio-Spatial & 63.5  & 82.0 & 46.6 & 48.1 & 66.0 &  33.5 & 56.3  & 79.4  & 41.7  & 48.8 & 70.2  & 35.6  \\
%     \textbf{Ours} &  \textbf{63.8} & \textbf{83.5} & \textbf{47.0} & \textbf{49.9} & 70.6 & 35.8 & \textbf{60.1} & \textbf{82.2} & \textbf{44.4} & \textbf{52.6 }& \textbf{72.5} & \textbf{37.9} \\
%     \bottomrule
%   \end{tabular}
%      \caption{Encoder ablation: \textit{w/o} refers to the model without the specified encoder. Results on the \textbf{static} Audio Goal task, trained with complex scenarios and evaluated \textbf{without} \& \textbf{with} complex scenarios on MP3D.} 
%   \label{tab:model_components}
% %   \vspace{-0.5cm}
% \end{table*}

\begin{table*}
  \vspace{-0.3cm}
 % \tablesize
  \centering
  \begin{tabular}{l|ccc|ccc}
    \toprule
    Model & \multicolumn{3}{c|}{Multiple Heard} & \multicolumn{3}{c}{Unheard} \\
    \cmidrule{2-7}
    & SPL & SR & SNA  & SPL & SR & SNA \\
    \midrule
     % w/o Depth &\textbf{ 64.9 } & \textbf{83.8  } & \textbf{47.8 } & \textbf{53.6 } & \textbf{72.9} & \textbf{38.8} & \textbf{60.1 } & 80.9  & 44.1 & \textbf{54.5 } & \textbf{ 74.7 } & \textbf{39.5 } & 1.2\\
     w/o Audio  & 55.6 & 78.8 & 41.5 & 49.6 & 72.1 & 37.0 \\
     w/o  Audio-Spatial & 56.3  & 79.4  & 41.7  & 48.8 & 70.2  & 35.6  \\
    \textbf{Ours} & \textbf{60.1} & \textbf{82.2} & \textbf{44.4} & \textbf{52.6 }& \textbf{72.5} & \textbf{37.9} \\
    \bottomrule
  \end{tabular}
     \caption{Encoder ablation: \textit{w/o} refers to the model without the specified encoder. Results on the \textbf{static} Audio Goal task, trained with complex scenarios and evaluated \textbf{with} complex scenarios on MP3D.} 
  \label{tab:model_components}
%   \vspace{-0.5cm}
\end{table*}
\begin{table*}
  \vspace{-0.3cm}
  \centering
  \begin{tabular}{l|ccc|ccc|ccc|ccc}
    \toprule
    Model & \multicolumn{6}{c|}{without complex} & \multicolumn{6}{c}{with complex} \\
    \cmidrule{2-13}
    & \multicolumn{3}{c|}{Multiple Heard} & \multicolumn{3}{c|}{Unheard} & \multicolumn{3}{c|}{Multiple Heard} & \multicolumn{3}{c}{Unheard}\\
    \cmidrule{2-13}
    & DSPL & SR & DSNA  & DSPL & SR & DSNA  & DSPL & SR & DSNA  & DSPL & SR & DSNA \\
    \midrule
    Ours + intersection reward & 54.8 & 82.5 & 30.9 & \textbf{34.1} & \textbf{54.8} & \textbf{19.1} & 52.5 & 91.5 & 31.2 & 33.8 & 57.6 & 20.2 \\
    \textbf{Ours} & \textbf{62.6} & \textbf{90.0} & \textbf{34.5} & 29.4 & 49.1 & 16.1 & \textbf{63.2} & \textbf{94.0} & \textbf{35.3} & \textbf{37.0}   & \textbf{57.7} & \textbf{21.0}\\
    \midrule
    Ours + comp + intersection reward & 39.6 & 64.3 & 22.4 & 29.9 & 55.0 & 16.9 & 63.1 & \textbf{94.3} & \textbf{36.9} & 37.3 & 63.6 & 22.0 \\
    \textbf{Ours + comp} & \textbf{43.9 }& \textbf{67.6} & \textbf{24.5} & \textbf{33.9} & \textbf{67.3} & \textbf{19.1 }&  \textbf{61.6} &  91.9 & 36.2 & \textbf{45.8} & \textbf{74.0} & \textbf{26.0}  \\ 

    \bottomrule
  \end{tabular}
    \caption{Reward ablation: \textit{+intersection reward} uses the distance to the shortest reachable intersection instead of the distance to the current position in the reward function. Results on the \textbf{dynamic} Audio Goal task \textbf{without} complex scenarios.  
  }
    \label{tab:dynamic_new_reward}

\end{table*}

\begin{table*}
  \vspace{-0.3cm}
  \centering
  \begin{tabular}{l|ccc|ccc|ccc|ccc}
    \toprule
    Model & \multicolumn{6}{c|}{without complex} & \multicolumn{6}{c}{with complex} \\
    \cmidrule{2-13}
    & \multicolumn{3}{c|}{Multiple Heard} & \multicolumn{3}{c|}{Unheard} & \multicolumn{3}{c|}{Multiple Heard} & \multicolumn{3}{c}{Unheard}\\
    \cmidrule{2-13}
    & DSPL & SR & DSNA  & DSPL & SR & DSNA  & DSPL & SR & DSNA  & DSPL & SR & DSNA \\
    \midrule

    Ours + comp + intersection reward & 32.1 & 54.2 & 18.1 & 28.4 & 48.5 & 16.0 & 55.1 & \textbf{92.8} & 31.9 & 43.8 & 77.2 & 25.9 \\
    \textbf{Ours + comp} & \textbf{38.6} & \textbf{63.8} & \textbf{21.8} & \textbf{32.9} &\textbf{58.9}  & \textbf{18.6} & \textbf{59.9} & \textbf{92.8} & \textbf{33.3} & \textbf{52.1} & \textbf{84.1}& \textbf{30.4}  \\ 

    \bottomrule
  \end{tabular}
    \caption{Reward ablation: \textit{+intersection reward} uses the distance to the shortest reachable intersection instead of the distance to the current position in the reward function. Results on the \textbf{dynamic} Audio Goal task \textbf{with} complex scenarios. 
  }
    \label{tab:dynamic_new_reward_complex}

\end{table*}

\section{Complexity of Audio Scenarios}

In this section, we further analyze the importance of the individual components of the audio scenarios. \cref{tab:decomp} shows the performance on the standard static AudioGoal task for models trained with subsets of these components. Focusing on unheard sounds, we find that training with the second audio improves performance on both datasets, while the spectrogram augmentations improve generalization on Replica, but not on Matterport3D. On the other hand, combining both further improves on Matterport3D but not on Replica. Lastly, the addition of distractor sounds further improves generalization on MP3D, but does not have a large impact on Replica. These differences in the impact of the individual elements across datasets show the importance to combine different randomizations and perturbations.

On the heard sounds, the perturbations do not influence performance on MP3D substantially. On Replica, the decomposition shows that the decrease was not due to any single component, but rather any kind of disturbance reduces the performance. This might further indicate that the agent learns to use or overly rely on some features of the audio signal that are not as robust or general to navigate these much smaller environments. On the other hand, the impact of the perturbations might be larger as the distances to the goals are generally shorter and thereby the number of audio observations the agent receives to filter out such disturbances is also fewer.

We then repeat the same evaluation with complex scenarios at test time, shown in \cref{tab:decomp_complex}. We find that training on all components is required to achieve robustness to these scenarios. No subset of the components is able to induce the same robustness to the full set of perturbations. In particular, the performance of the model trained on clean audio scenarios drops significantly across all settings and metrics, often more than 50\%. This demonstrates the importance of training on these scenarios to achieve robust behavior in noisy and complex audio environments.

% We find that on Replica all subsets improve performance on unheard sounds. While the results on Replica heard show that the decrease was not due to any single component. On Matterport3D unheard sounds large gains occur as soon as we combine at least two of the components, with the best SPL for the combination of all of them. These differences between the kind of disturbance indicate the importance to combine different randomizations and perturbations.
% We then repeat the same evaluation with complex scenarios at test time, shown in \cref{tab:decomp_complex}. Here we very clearly find that training on all components is required to achieve robustness to these scenarios.

\section{Motion Model}
To investigate the effect of the speed of the moving sounds, we test the agent trained on static sound sources plus sound sources with a movement probability of $p=30\%$ on unseen $p \in \{10\%, 20\%, 30\%, 40\%\}$. The results are shown in \cref{tab:moving_percentage}. The agent generalizes well to these unseen motion patterns, with success rates staying in a close range of 80.5\%-88.9\% and 89.8\%-93.6\% on Replica and MP3D heard respectively. It actually performed better on lower, but unseen, values of $p$, further demonstrating that the agent does not overfit a motion pattern. 

\section{Hyperparameters}

To ensure a fair comparison to the baselines we use the same hyperparameters as reported by \cite{chen2020learning} for all models, including our proposed architecture. The hyperparameters are listed in \cref{tab:hyper}. Each episode has a time limit of 500 steps after which it will be stopped and counted as a failure.

\myworries{\section{Audio Encoder Ablation}\label{sec:model_components_supp}
To evaluate the importance of the architecture, we compare to the same architecture without either of the audio encoders. \cref{tab:model_components} reports the results on the MP3D dataset. We find both the audio and the audio-spatial encoder to be important for performance on both heard and unheard sounds.}

\myworries{\section{Dynamic Sound Reward}\label{sec:dynamic_sound_reward}
As we discuss in \cref{sec:reward}, for dynamic sounds the dense distance reward no longer directly points to the optimal policy, reducing the value of the supervisory signal. While we could replace it with the distance to the earliest possible intersection, this point depends on the sound target's movements and as such is a priory unknown to the agent. As a consequene, this reward may appear stochastic from the perspective of the agent, potentially making the value estimation and optimization problem more difficult. We evaluate the effect of replacing the distance reward with this intersection reward in \cref{tab:dynamic_new_reward} and \cref{tab:dynamic_new_reward_complex}. The results support the hypothesis that the task for dynamic sounds is not necessarily easier to solve with this reward, with average success rates being higher with original "static" distance reward.}

% \comm{Things mentioned in the paper that we'll do:

% \begin{itemize}
    % \item Exact architecture (layers of each module)
    % \item Visualisation of the behaviour on the dynamic task
    % \item Decomposition of the effects of the individual elements of the complex scenarios
    % \item Hyperparameters for ours (and maybe baselines)
    % \item (Anonymized) code
    % \item video
% \end{itemize}}

% \comm{
% Possible things to add here:
% \begin{itemize}
%     \item Comparison with noisy evaluation as done in \cite{chen2020learning}
%     \item Influence of leaving out each three inputs to the GRU: audio, audio-visual and depth. Also without the change in action parametrization.
%     \item Trained on non-complex and evaluated on complex to show that these are difficult scenarios?
%     \item Second metric: Naive agent that always takes step towards current sound source position (in contrast to DSPL that compares with movement towards closest reachable intersection)
%     \item Evaluating AV-WaN with chosing waypoints each env-step
% \end{itemize}

% \input{tables/baseline_reproduction}

% }

% {\small
% \bibliographystyle{splncs04}
% \bibliography{egbib}
% }
% \end{document}

\end{document}